\documentclass{jpsj2}
\usepackage{dcolumn} 
 \usepackage{color} 
 \usepackage{graphicx} 
 \usepackage{amsmath,amsthm,amssymb} 
 \usepackage{bm}

\newcommand{\Ev}{{\bm E}}

\newcommand{\ef}{\epsilon_{\rm F}} 
\newcommand{\ek}{\epsilon_{\bm k}}

\newcommand{\Jex}{{J_{\rm ex}}}
 
\newcommand{\jsv}{{\jv_{\rm s}}} 
\newcommand{\jv}{{\bm j}}

\newcommand{\gr}{g^{\rm r}} 
\newcommand{\ga}{g^{\rm a}} 
\newcommand{\kv}{{\bm k}} 
\newcommand{\pv}{{\bm p}} 
\newcommand{\kf}{k_{\rm F}} 

\newcommand{\nimp}{{n_{\rm imp}}} 
\newcommand{\imp}{\rm imp} 
\newcommand{\qv}{{\bm q}}
\newcommand{\Qv}{{\bm Q}} 
 
\newcommand{\sigmam}{\bm \sigma}  
\newcommand{\Sv}{{\bm S}} 
\newcommand{\tr}{{\rm tr}}
\newcommand{\xv}{{\bm x}}

\newcommand{\lamso}{{ \lambda_{\rm{so}}}} 
\newcommand{\la}{\langle} 
\newcommand{\ra}{\rangle} 

\newcommand{\Av}{\bm{A}} 
\newcommand{\nv}{\bm{n}}

\title{Perturbation theory of the dynamic inverse spin Hall effect with charge conservation}

\author{Kazuhiro Hosono, Akihito Takeuchi, and Gen Tatara}

\inst{Department of Physics, Tokyo Metropolitan University, Hachioji, Tokyo 192-0397, Japan}
\abst{We present gauge-invariant theory of the dynamic inverse spin Hall effect driven by the spin--orbit interaction in metallic systems.
Charge conservation is imposed diagrammatically by including vertex corrections. 
We show the charge current is induced by an effective electric field that is proportional to the spin current pumped by the magnetization dynamics. 
The result is consistent with recent experiments.
}
\kword{spintronics, inverse spin Hall effect, spin pumping, spin current, spin Hall effect, spin--orbit interaction, Keldysh formalism}

\begin{document}
\maketitle
\section{Introduction} 
Control and manipulation of spin current, the flow of the spin degree of freedom in solids, has been extensively studied since the discovery of the spin Hall effect 
in semiconductor systems \cite{Kato,JW,Murakami,Shinova}. The spin Hall effect converts electric voltage into spin current via the relativistic spin--orbit interaction. The induced spin current was first detected indirectly by observing spin accumulation at the edge of the sample \cite{Kato,JW}. The spin--orbit interaction also works in the opposite way; that is, when the spin--orbit interaction acts on the spin current, charge current is expect to arise. This is the inverse spin Hall effect \cite{Saitoh06,Takahashi02,Takahashi08}. By use of the inverse spin Hall effect, direct electrical detection of the spin current has become possible \cite{Saitoh06,Ando, Kimura07,Valenzuela06,Seki}.

Depending on the creation mechanism of the spin current, there are two inverse spin Hall effects. 
One is the static inverse spin Hall effect involving non-local spin injection \cite{Kimura07,Valenzuela06,Seki}. Spin current is created from the electric field via the direct spin Hall effect and current arising from the inverse spin Hall effect can be observed by attaching a nonmagnetic contact
of Pt or Au having strong spin--orbit interaction. The other effect is the dynamic inverse spin Hall effect involving a spin pumping mechanism \cite{Brataas00,Tserkovnyak02,Wang06} with dynamic magnetization.
The concept was employed by Saitoh who used a microwave to excite the uniform precession of the magnetization. The resulting electric voltage was found to be
perpendicular to the spin current \cite{Saitoh06,Ando}. 

The creation of electric current by dynamic magnetization is well known in classical electromagnetism as Faraday's induction law.  
In contrast to Faraday's law, which is classical and macroscopic, the inverse spin Hall effect is a quantum relativistic effect that is induced in solids. The crucial practical difference between the two is that while the current induced by Faraday's law diminishes when the system size is reduced with the field kept constant, the efficiency of the inverse spin Hall mechanism remains constant. The use of the dynamic inverse spin Hall effect is therefore expected to be essential in high-density integrated devices.

Theoretical study of non-local spin injection was carried out by Takahashi and Maekawa \cite{Takahashi02,Takahashi08}. Their discussion was based on a semiclassical
argument that the charge current and spin current are related by the spin--orbit interaction $\jv\propto \sigmam\times\jsv$, where $\sigmam$ and $\jsv$ 
represent the spin polarization and flow direction of the spin current, respectively. Their work qualitatively explained the experimentally observed 
static inverse spin Hall effect \cite{Kimura07,Valenzuela06,Seki}. However, the theory, which simply predicts the generation of charge current when spin current and a spin--orbit interaction exist, violates the conservation law for charge. Charge conservation is a crucial issue in any physical phenomenon and a framework that upholds the conservation law is certainly required.

The dynamic inverse spin Hall effect was studied theoretically for two-dimensional disordered electron systems with Rashba interaction \cite{Ohe07,Takeuchi08}.
Coupling to the localized spins was treated perturbatively assuming weak coupling as is the case for ferromagnet/normal metal junctions. The dominant contribution to
the current was shown to be $j_\alpha\simeq \lambda_{\rm R} \epsilon_{\alpha\beta z} \langle (\Sv\times\dot{\Sv})^\beta \rangle$, where $\Sv$ denotes a local spin. $\lambda_{\rm R}$ and $\epsilon_{\alpha \beta z}$ are the strength of the Rashba interaction and anti-symmetric tensor respectively, $\la\ \ra$ represents the average owing to electron's diffusive motion and the $z$ axis is perpendicular to the plane of the junction. The result seems reasonable since the current arises from the energy dissipation (damping) of the spin system, $\Sv\times\dot{\Sv}$. The obtained current direction is
consistent with experimental results \cite{Saitoh06,Ando}. It is, however, an open and critical issue whether the Rashba inverse spin Hall mechanism applies to the experimental systems \cite{Saitoh06,Ando}, since the Rashba interaction known so far in metallic systems occurs on the surface of nonmagnetic metals \cite{Ast07} and not at the interface between a ferromagnet and nonmagnet.

Gauge-invariant (charge-conserved) theory for the dynamic inverse spin Hall effect in metallic (i.e., symmetric) spin--orbit systems is therefore necessary.
The aim of this paper is to present such theory  
 on the basis of microscopic quantum many-body theory. Charge conservation is guaranteed by evaluating Feynman diagrams while keeping the gauge invariance. 
 This means we include vertex corrections, which represent the effect of electron diffusion. 
As the spin--orbit interaction, we consider that induced by random impurities, which is dominant in metallic systems.
A crucial feature of the present system is that the translational and rotational invariances are recovered after averaging over the random impurities, in sharp contrast to the case for a Rashba system with explicitly broken symmetry \cite{Ohe07,Takeuchi08}.

We will show that the induced current is written as
\begin{align}
j_\alpha = \alpha_{\rm ISH}\epsilon_{\alpha \beta \gamma}j_{{\rm s},\beta}^{\gamma}-D\partial_\alpha \rho,
\label{eq:jfull} 
\end{align}
where $j_{{\rm s},\beta}^{\gamma}$ is the spin current density locally created by the dynamic magnetization,
$\alpha_{\rm ISH}$ represents the efficiency of the spin--charge conversion, $D\equiv \frac{2\ef \tau}{3m}$ is the diffusion constant and $\rho$ is the charge density. ($\ef$ and $\tau$ the Fermi energy and the lifetime, respectively) The coefficient $\alpha_{\rm ISH}$ and expressions for  $j_{{\rm s},\beta}^{\gamma}$ and $\rho$ are given below. eq. (\ref{eq:jfull}) is exact within second order perturbation theory and for slowly varying spin structures. The result satisfies the charge conservation law $\dot{\rho}+\nabla\cdot\jv=0$.
The first term of eq. (\ref{eq:jfull}) clearly indicates that spin--charge conversion indeed occurs in dynamic inverse spin Hall systems. 
This spin-pumped part is local and thus the spin current acts as an effective electric field. 
In contrast, the spin current in static inverse spin Hall systems arises diffusively from spin accumulation 
\cite{Takahashi02,Takahashi08} and thus the static inverse spin Hall current is expressed by a gradient term $\nabla s$, where $s$ is the electron spin density.
In the present dynamic case, diffusive spin current, proportional to $\nabla s$, does not exist.
The physical origins of the spin currents for the dynamic and static inverse spin Hall effects are therefore different.
\section{Model}
The system we consider is a disordered electron system with lifetime $\tau$ in the presence of a random spin--orbit interaction and coupled to dynamic localized spin $\Sv(\xv,t)$. 
The total Hamiltonian is given by 
\begin{align*}
H =-\frac{\hbar^2}{2m}\int{d^{3}x}\psi(\xv,t)^\dagger\nabla^2 \psi(\xv,t)+ H_{\rm imp}+H_{\rm ex}+H_{\rm so},
\end{align*}
where $\psi^\dagger$ and $\psi$ are the electron creation and annihilation operators.
The first term describes conduction electron and $H_{\rm imp}=\int{d^{3}x}\psi(\xv,t)^\dagger U(\xv)\psi(\xv,t)$
represents elastic impurity scattering, where $U(\xv)$ is the random impurity potential. We assume averaging over the impurity position is carried out in the standard manner as $ \la U(\xv)U(\xv') \ra_{\imp}=u_0^2\nimp\delta(\xv-\xv')$, where $u_0$ and $\nimp$ are the strength of the impurity potential and the impurity concentration, respectively. 
This impurity scattering $ H_{\rm imp}$ gives lifetime as $\tau=\frac{\hbar V}{2\pi N(0) n_{\rm imp}u_{0}^2}$, where $N(0)\equiv \frac{mV\kf}{2\pi^2\hbar^2}$ is electron's density of states at the Fermi energy and $\it V$ is the system volume. The electron's diffusion motion expressed by the diffusion ladder also arises from the impurity scattering.
 The coupling to the localized spin is represented by the $s$--$d$ type exchange interaction, 
\begin{align*} 
H_{\rm ex}=J_{\rm ex}\int{d^{3}x}
\Sv(\xv,t)\cdot\psi^{\dagger}(\xv,t)\sigmam\psi(\xv,t).
\end{align*}
 We assume this
interaction is weak (compared with the inverse lifetime), $\frac {\Jex\tau}{\hbar} \ll1$. In the case of ferromagnet--nonmagnet junctions, the electrons contributing to the
measured current are those in the nonmagnet, and the exchange interaction acts only at the interface. The assumption of a weak exchange interaction would thus be correct. The
spin--orbit interaction we consider is due to random impurities \cite{Dugaev1,WK,KAM}: 
\begin{align*} 
H_{\rm so}=-\frac{i\lamso}{2}\epsilon_{i j k}\int{d^{3}x} \partial_{i}U(\xv)
\left(\psi^{\dagger}(\xv,t)\sigma^k \overleftrightarrow{\partial_j}\psi(\xv,t)\right), 
\end{align*} 
where $\psi^{\dagger}\overleftrightarrow{\partial_{j}}\psi=\psi^{\dagger}{\partial_{j}}\psi-({\partial_{j}}\psi^{\dagger})\psi$, $\lamso$ represents the spin--orbit strength and $U(\xv)$ is same impurity potential as in $H_{\rm imp}$. The spin--orbit interaction is treated to the first order.

\section{Method}
Let us calculate the electric current induced from the magnetization dynamics.
We assume a slowly varying spin structure in both space and time, and expand the current with respect to $\partial_t$ and $\nabla$. The total electric current is given as $j_\alpha=\tilde{j}_\alpha + j^{\rm SO}_\alpha$, where
\begin{align*}
&\tilde{j}_\alpha = -i\frac{e\hbar}{2m}\langle\psi^{\dagger}\overleftrightarrow{\partial_\alpha}\psi\rangle,
\\
&j^{\rm SO}_\alpha = \frac{e\lamso}{\hbar}\epsilon_{\alpha l m}\langle\psi^{\dagger}(\sigma^l\partial_{m} U)\psi\rangle,
\end{align*}
 and the charge density as $\rho\equiv e\la\psi^{\dagger}\psi\ra$.
Here, $\tilde{j}_\alpha$ is the normal term (skew--scattering term) and $ j^{\rm SO}_\alpha$ is the anomalous term which arises from the spin--orbit interaction due to the impurity (side jump term).
The expectation values are written by the lesser Green's function as  
 \begin{align*}
&\tilde{j}_{\alpha}(\xv,t)=-i\frac{e\hbar^2}{mV}\sum_{\kv,\qv}k_{\alpha} \lim_{t'\to t} \la \tr G^{<}_{\kv-\frac{\qv}{2},\kv+\frac{\qv}{2}}(t,t')\ra_{\imp} e^{ -i\qv\cdot \xv},
\\
&j^{\rm SO}_{\alpha}(\xv,t)=\frac{e\lambda_{so}}{V^2}\sum_{\kv \kv' \qv}\epsilon_{ilm}\delta_{m \alpha}\left(k'-k\right)^{l}\lim_{t'\to t}\la V_{\kv'-\kv}\tr[\sigma^{i}G^{<}_{\kv-\frac{\qv}{2},\kv'+\frac{\qv}{2}}(t,t')]\ra_{\imp} e^{ -i\qv \cdot \xv}
\\
&\rho(\xv,t)=-i\frac{e\hbar}{V} \sum_{\kv,\qv}\la \tr G^{<}_{\kv-\frac{\qv}{2},\kv+\frac{\qv}{2}}(t,t')\ra_{\imp} e^{ -i{\qv\cdot \xv}}.
\end{align*}
Here, trace ($\tr$) is over spin, and $\langle\ldots\rangle _{\rm{imp}}$ represents the impurity averaging. 
Lesser Green's function $G^{<}_{\kv,\kv'}(t,t')$ is calculated by use of the contour ordered Green's function $G_{\kv,\kv'}(t,t')=\frac{1}{i\hbar}\la T_c[c_{\kv}(t) c^{\dagger}_{\kv'}(t')]\ra,$
where $T_c$ is the contour-ordering operator along the contour $C$ of the complex time \cite{Hung}. The lesser Green's function satisfies the Dyson given as 
\begin{align*}
G^{<}_{\kv,\kv'}(t,t')=&g^{<}_{\kv}(t-t')\delta_{\kv,\kv'}\\
&+J_{\rm ex}\sum_{\pv, i}\left(\int_c dt_{1} g_{\kv}(t-t_{1})S^{i}_{\pv-\kv}(t_{1})\sigma^{i}G_{\pv,\kv'}(t_{1},t')\right)^<
\\
&+\frac{i\lamso}{V}\sum_{\pv, i}U_{\kv-\pv}(\kv \times \pv)_{i} \left(\int_c dt_{1}g_{\kv}(t-t_{1})\sigma^{i}G_{\pv,\kv'}(t_{1},t')\right)^<,
\end{align*}
where $g_{\kv}(t-t')$ is the free Green's function including the lifetime.
The Fourier component of the impurity potential
is defined as $U_{\kv'-\kv}=\int d\xv U(\xv)e^{-i(\kv-\kv')\cdot \xv}$.
In the Fourier space (defined as $g_{\kv}(t-t')=\int\frac{d\omega}{2\pi}g_{\kv,\omega}e^{-i\omega(t-t')}$), the free lesser component is obtained as 
$g^<_{\kv,\omega}=f(\omega)(\ga_{\kv,\omega}-\gr_{\kv,\omega})$, where $f(\omega)$ is the Fermi distribution function, and $\gr_{\kv,\omega}$ and  $\ga_{\kv,\omega}=(g^r_{\kv,\omega})^\ast$ are the retarded and the advance Green's function respectively. 
The retarded Green's function is given as $\gr_{\kv,\omega}=\frac{1}{\hbar \omega-(\ek-\ef) + i\eta}$ ($\eta\equiv \frac{\hbar}{2\tau}$). (The free Green's functions are diagonal in spin space.) 
\subsection{Linear order in $\Jex$}
\begin{figure} 
\begin{center}
\resizebox{9cm}{!}{\includegraphics[angle=0]{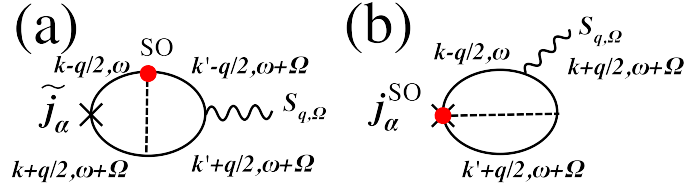}}
 \caption{Diagrammatic representation of the charge current at linear order in the exchange interaction.
 Crosses and filled circles represent the current ($\tilde{j}$) and spin--orbit interaction (SO), respectively.
The intersection with a filled circle represents the contribution of the anomalous velocity ($j^{\rm SO}$) from the spin--orbit interaction. 
 Solid and wavy lines denote the impurity-averaged electron Green's function and the exchange interaction $\Sv$, respectively. Dashed lines denote the impurity averaging between the spin--orbit interaction and the impurity scattering. Complex conjugate to each process (a) and (b) are also included in eq. (\ref{eq:current1}).}
 \end{center}
  \end{figure}

The contributions to the charge current at the linear order in both $\Jex$ and $\lamso$ are shown in Fig. 1. These contributions arise from the second order in impurity potential $U_{\kv-\kv'}$ (One of them is arising from $H_{\rm so}$, the other one from $H_{\rm imp}$).  
 The vertex corrections to the normal charge current at the linear order in $\Jex$ vanish exactly.
 The charge current at the linear order in the exchange coupling $\Jex$ is thus obtained as
\begin{align}
&{j}^{(1)}_{\alpha}(\qv,\Omega)=
\frac{2e\Jex \lamso u_{0}^2 \nimp}{V^2}\epsilon_{\gamma j\beta}S^\gamma_{\qv,\Omega}
\left({\cal C}^{1(a)}_{\alpha j \beta; \qv,\Omega} +{\cal C} ^{1(b)}_{\alpha j \beta; \qv,\Omega}\right),
\label{eq:current1}
\end{align}
where  $\Sv_{\qv}(\Omega)$ is Fourier component of the localized spin, defined as $\Sv(\xv,t)=\sum_{\qv,\Omega}\Sv_{\qv}(\Omega)e^{i(\Omega t-\qv\cdot \xv)}$.
Contribution from an each diagram (a) and (b) of Fig. 1 are given as
\begin{align*}
{\cal C}^{1(a)}_{\alpha j \beta; \qv,\Omega}=&\frac{\hbar^2}{m}\sum_{\kv \kv'} k^{\mu} \left(k'-k\right)^j
q^\beta\left(\int\frac{d\omega}{2\pi}g_{\kv-\frac{\qv}{2},\omega}g_{\kv'-\frac{\qv}{2},\omega}g_{\kv'+\frac{\qv}{2},\omega +\Omega}g_{\kv+\frac{\qv}{2},\omega +\Omega}\right)^<,
\\
{\cal C}^{1(b)}_{\alpha j \beta; \qv,\Omega}=&-\delta_{j\alpha}\sum_{\kv,\kv'}\left(k'-k\right)^\beta
\\
&\times
\left[\left(\int\frac{d\omega}{2\pi}g_{\kv-\frac{\qv}{2},\omega}g_{\kv'-\frac{\qv}{2},\omega}g_{\kv'+\frac{\qv}{2},\omega +\Omega}\right)^<
+\left(\int\frac{d\omega}{2\pi}g_{\kv-\frac{\qv}{2},\omega}g_{\kv+\frac{\qv}{2},\omega+\Omega}g_{\kv'+\frac{\qv}{2},\omega +\Omega}\right)^<\right].
\end{align*}
${\cal C}^{1(a)}_{\alpha j \beta; \qv,\Omega}$ is the contribution of the normal charge current $(\tilde{j}_{\alpha}$), and ${\cal C}^{1(b)}_{\alpha j \beta; \qv,\Omega}$ is arising from the additional charge current ($j^{\rm SO}_{\alpha}$).
Using Langreth theorem \cite{Hung}, we obtain
\begin{align}
\left(\int\frac{d\omega}{2\pi}g_{\kv-\frac{\qv}{2},\omega}g_{\kv'-\frac{\qv}{2},\omega}g_{\kv'+\frac{\qv}{2},\omega +\Omega}g_{\kv+\frac{\qv}{2},\omega +\Omega}\right)^<
 =&\int\frac{d\omega}{2\pi}\left[-f(\omega+\Omega)\gr_{\kv-\frac{\qv}{2},\omega}\gr_{\kv'-\frac{\qv}{2},\omega}\gr_{\kv'+\frac{\qv}{2},\omega+\Omega}\gr_{\kv+\frac{\qv}{2},\omega+\Omega}\right.
\nonumber
\\
 &+\left(f(\omega+\Omega)-f(\omega)\right)\gr_{\kv-\frac{\qv}{2},\omega}\gr_{\kv'-\frac{\qv}{2},\omega}\ga_{\kv'+\frac{\qv}{2},\omega+\Omega}\ga_{\kv+\frac{\qv}{2},\omega+\Omega}
\nonumber
\\
&+\left.f(\omega)\ga_{\kv-\frac{\qv}{2},\omega}\ga_{\kv'-\frac{\qv}{2},\omega}\ga_{\kv'+\frac{\qv}{2},\omega+\Omega}\ga_{\kv+\frac{\qv}{2},\omega+\Omega}\right].
\label{eq:Lang}
\end{align}
Since the contribution containing only $\gr$ or $\ga$ only is small (higher order of $\frac{\hbar}{\ef\tau}$) after integrating over $\kv$ and $\kv'$,  the dominant contribution of ${\cal C}^{1(a)}_{\alpha j \beta; \qv,\Omega}$ come from the second term of the right-hand side of eq. (\ref{eq:Lang}). By using the approximation $f(\omega+\Omega)-f(\omega)\simeq\Omega f'(\omega)$ (justified if $\Omega \ll \omega$) and $f'(\omega)\simeq -\delta(\omega)$ at low temperature, we obtain 
\begin{align*}
{\cal C}^{1(a)}_{\alpha j \beta; \qv,\Omega} \simeq -\frac{\Omega}{2\pi}\frac{\hbar^2}{m}\sum_{\kv'}k^{\alpha}\left(k'-k\right)^j q^\beta\gr_{\kv-\frac{\qv}{2}}\gr_{\kv'-\frac{\qv}{2}}\ga_{\kv'+\frac{\qv}{2}}\ga_{\kv+\frac{\qv}{2}},
\end{align*}
where $\gr_{\kv}=\gr_{\kv,\omega=0}$ etc.
The other contribution ${\cal C}^{1(b)}_{\alpha j \beta; \qv,\Omega}$ is similarly obtained as
\begin{align*}
{\cal C}^{1(b)}_{\alpha j \beta; \qv,\Omega} &\simeq \frac{\Omega}{2\pi}\delta_{j\alpha}\sum_{\kv,\kv'}\left(k'-k\right)^\beta\left(\gr_{\kv-\frac{\qv}{2}}\gr_{\kv'-\frac{\qv}{2}}\ga_{\kv'+\frac{\qv}{2}}
+\gr_{\kv-\frac{\qv}{2}}\ga_{\kv+\frac{\qv}{2}}\ga_{\kv'+\frac{\qv}{2}}\right)
\\
&= \frac{\Omega}{2\pi}\delta_{j\alpha}\frac{\hbar^2}{m}\sum_{\kv,\kv'}k'^\beta(\kv'-\kv)\cdot\qv \gr_{\kv-\frac{\qv}{2}}\gr_{\kv'-\frac{\qv}{2}}\ga_{\kv'+\frac{\qv}{2}}\ga_{\kv+\frac{\qv}{2}}
\end{align*}
Here, we used the identity $\gr_{\kv-\frac{\qv}{2}}-\ga_{\kv+\frac{\qv}{2}}=\left(\epsilon_{\kv-\frac{\qv}{2}}-\epsilon_{\kv+\frac{\qv}{2}} -2i\eta\right)\gr_{\kv-\frac{\qv}{2}}g^a_{\kv+\frac{\qv}{2}}$.
We assume the smooth magnetization profile ($|\qv|l\ll1$, where $l$ is the electron mean free path), and expand ${\cal C}^{1(a)}_{\alpha j \beta; \qv,\Omega}$ and ${\cal C}^{1(b)}_{\alpha j \beta; \qv,\Omega}$ with respect to wave vector $\qv$. We also assume that the system is spatially symmetric. We then obtain
\begin{align}
{\cal C}^{1(a)}_{\alpha j \beta; \qv,\Omega}={\cal C}^{1(b)}_{\alpha j \beta}(\qv,\Omega)=\frac{\Omega}{3\pi}\delta_{\alpha j}q^\beta \sum_{\kv \kv'}\epsilon_{\kv}|\gr_{\kv}|^2|\ga_{\kv'}|^2.
\end{align}
Therefore the contribution of the skew--scattering and side jump are same in the slowly varying limit and at the linear order in $\Jex$.
Integration over $\kv$, $\kv'$ is carried out using $\sum_{\kv}\epsilon_{\kv}|\gr_{\kv}|^2=N(0)\pi\frac{\ef}{\eta},  \sum_{\kv'}|\gr_{\kv'}|^2=\frac{N(0)\pi}{\eta}$. We finally obtain the current eq. (2) as 
\begin{align}
{j}^{(1)}_{\alpha}(\qv,\Omega)=\alpha_{\rm ISH}c\epsilon_{\alpha\beta \gamma}\Omega q^\beta  S^\gamma_{\qv}(\Omega),
\label{eq:reslutj1}
\end{align}
where $c\equiv -\frac{4e\Jex N(0)\ef \tau^2}{3mV}$ and $\alpha_{\rm ISH}\equiv -\frac{\hbar\lamso \kf^2 }{\ef \tau}$.
\subsection{Second order in $\Jex$}
We next calculate the contribution at the second order in the exchange coupling $\Jex$. Total charge current at the second order in $\Jex$ is given as
\begin{align}
j^{(2)}_\alpha(\qv, \Qv, \Omega_1, \Omega_2)=
i\frac{2e\Jex^2\lamso u_0^2 n_{\imp}}{V^2}\epsilon_{ijk}S^j_{\frac{\Qv+\qv}{2},\Omega_1}S^k_{\frac{\Qv-\qv}{2},\Omega_2}\left({\cal D}^{(2)}_{\alpha i;\qv,\Qv,\Omega_1,\Omega_2}+{\cal D}^{(3)}_{\alpha i; \qv,\Qv,\Omega_1,\Omega_2}\right).
\label{eq:j2}
\end{align}
Here the contribution of ${\cal D}^{(2)}_{\alpha i;\qv,\Qv,\Omega_1,\Omega_2}$ is the local contribution which does not contain the vertex corrections as diagrammatically shown in Fig. 2. 
The diffusive contribution of the charge current is presented by ${\cal D}^{(3)}_{\alpha i;\qv,\Qv,\Omega_1,\Omega_2}$, which contains the vertex corrections to $\tilde{j}_{\alpha}$, and are shown in Fig. 3. 

\begin{figure} 
\begin{center}
\resizebox{11cm}{!}{\includegraphics[angle=0]{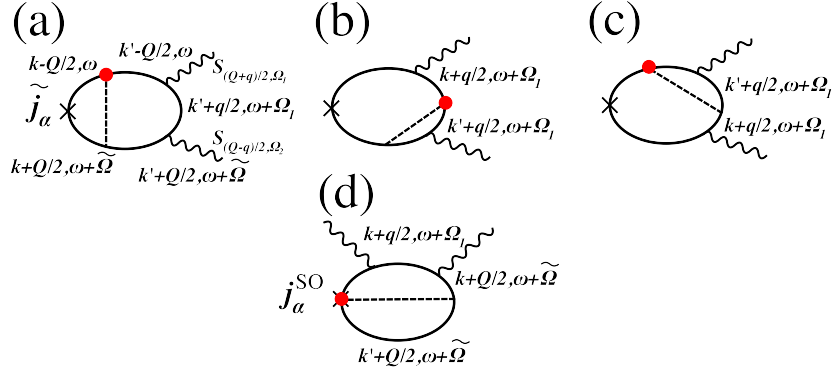}}
 \caption{Diagrammatic representation of the charge current at the second order in the exchange interaction. Complex conjugate processes are also included in eq. (\ref{eq:j2}).
}
 \end{center}
  \end{figure}
\subsubsection{Local contribution}
We calculate first the local contribution of the charge current.  
This contribution is given as 
${\cal D}^{(2)}_{\alpha i;\qv,\Qv,\Omega_1,\Omega_2}\equiv{\cal D}^{2(a)}_{\alpha i;\qv,\Qv,\Omega_1,\Omega_2}+{\cal D}^{2(b+c)}_{\alpha i;\qv,\Qv,\Omega_1,\Omega_2}+{\cal D}^{2(d)}_{\alpha i;\qv,\Qv,\Omega_1,\Omega_2}$, where
\begin{align*}
{\cal D}^{2(a)}_{\alpha i;\qv,\Qv,\Omega_1,\Omega_2}&=\frac{\hbar^2}{m}\sum_{\kv,\kv'}k^\alpha \left[\left(\kv-\frac{\Qv}{2}\right) \times \left(\kv'-\frac{\Qv}{2}\right)+\left(\kv'+\frac{\Qv}{2}\right) \times \left(\kv+\frac{\Qv}{2}\right)\right]^i
\\
&\times \left(\int\frac{d\omega}{2\pi}g_{\kv-\frac{\Qv}{2},\omega}g_{\kv'-\frac{\Qv}{2},\omega}g_{\kv'+\frac{\qv}{2},\omega +\Omega_1}g_{\kv'+\frac{\Qv}{2},\omega +\tilde{\Omega}}
 g_{\kv+\frac{\Qv}{2},\omega +\tilde{\Omega}}\right)^<,
 \\
{\cal D}^{2(b+c)}_{\alpha i;\qv,\Qv,\Omega_1,\Omega_2}&=\frac{\hbar^2}{m}\sum_{\kv,\kv'}k^\alpha \left[\left(\kv'+\frac{\Qv}{2}\right) \times \left(\kv+\frac{\Qv}{2}\right)-\left(\kv+\frac{\qv}{2}\right) \times \left(\kv'+\frac{\qv}{2}\right)\right]^i
\\
&\times \left(\int\frac{d\omega}{2\pi}g_{\kv-\frac{\Qv}{2},\omega}g_{\kv+\frac{\qv}{2},\omega+\Omega_1}g_{\kv'+\frac{\qv}{2},\omega +\Omega_1}g_{\kv'+\frac{\Qv}{2},\omega +\tilde{\Omega}}
 g_{\mathbf{k+\frac{Q}{2}},\omega +\tilde{\Omega}}\right)^<
 \\
&+\frac{\hbar^2}{m}\sum_{\kv,\kv'}k^\alpha \left[\left(\kv-\frac{\Qv}{2}\right) \times \left(\kv'-\frac{\Qv}{2}\right)-\left(\kv'+\frac{\qv}{2}\right) \times \left(\kv+\frac{\qv}{2}\right)\right]^i
\\
&\times \left(\int\frac{d\omega}{2\pi}g_{\kv-\frac{\Qv}{2},\omega}g_{\kv'-\frac{\Qv}{2},\omega}g_{\kv'+\frac{\qv}{2},\omega +\Omega_1}g_{\kv'+\frac{\qv}{2},\omega +\Omega_1}
 g_{\mathbf{k+\frac{Q}{2}},\omega +\tilde{\Omega}}\right)^<,
 \\
 {\cal D}^{2(d)}_{\alpha i;\qv,\Qv,\Omega_1,\Omega_2}&=
\epsilon_{ilm}\delta_{\alpha m}\sum_{\kv,\kv'}\left(k'-k\right)^l
\left[
\left(\int\frac{d\omega}{2\pi}g_{\kv-\frac{\Qv}{2},\omega}g_{\kv+\frac{\qv}{2},\omega+\Omega1}g_{\kv+\frac{\Qv}{2},\omega+\tilde{\Omega}}g_{\kv'+\frac{\Qv}{2},\omega+\tilde{\Omega}}\right)^<
\right.
\\
&\left.
+\left(\int\frac{d\omega}{2\pi}g_{\kv-\frac{\Qv}{2},\omega}g_{\kv'-\frac{\Qv}{2},\omega} g_{\kv'+\frac{\qv}{2},\omega+\Omega1} 
g_{\kv'+\frac{\Qv}{2},\omega+\tilde{\Omega}}\right)^<
\right],
\end{align*}
where $\tilde{\Omega}=\Omega_1+\Omega_2$. Each contribution corresponds to the contribution labelled in Fig. 2. After taking the lesser component, we obtain the leading contributions as 
  \begin{figure} 
\begin{center}
\resizebox{10cm}{!}{\includegraphics[angle=0]{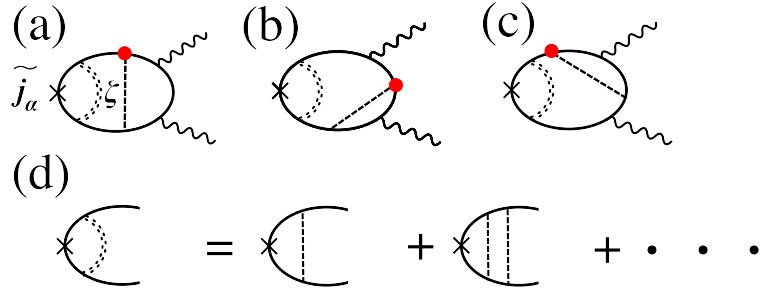}}
 \caption{Diagrammatic representation of the charge current with vertex correction due to the electron diffusion ladder ($\zeta$) shown by double dotted lines (shown in (a),(b),(c)). There are two diagrams of each type (a), (b) and (c) due to the complex conjugate process. 
 The diffusion ladder $\zeta$ ( shown in (d) ) describes the diffusive electron motion and results in non-local contributions, which sum to be $D\partial_{\alpha}\rho$. These processes containing $\zeta$ are the vertex corrections that are essential for guaranteeing the charge conservation.
 }
 \end{center}
  \end{figure}

\begin{align*}
{\cal D}^{2(a)}_{\alpha i;\qv,\Qv,\Omega_1,\Omega_2}
&=-\frac{\hbar^2}{2\pi m}\epsilon_{ilm}\Omega_1\sum_{\kv,\kv'}k^\alpha \left(k'-k\right)^lQ^m  
\gr_{\kv-\frac{\Qv}{2}}\gr_{\kv'-\frac{\Qv}{2}}\ga_{\kv'+\frac{\Qv}{2}}\ga_{\kv+\frac{\Qv}{2}}(\ga_{\kv'+\frac{\qv}{2}}-\gr_{\kv'+\frac{\qv}{2}}),
\\
{\cal D}^{2(b+c)}_{\alpha i;\qv,\Qv,\Omega_1,\Omega_2}&=
\frac{\hbar^2}{2\pi m}\Omega_{1}\epsilon_{ilm}\sum_{\kv,\kv'}
 k^\alpha\left(2k'^lk^m+\left(k'-k\right)^l(\frac{Q-q}{2})^m\right)\left(\gr_{\kv-\frac{\Qv}{2}}\gr_{\kv-\frac{\qv}{2}}\gr_{\kv'-\frac{\qv}{2}}\ga_{\kv'+\frac{\Qv}{2}}\ga_{\kv+\frac{\Qv}{2}}-\rm{c.c}\right),
 \\
 {\cal D}^{2(d)}_{\alpha i;\qv,\Qv,\Omega_1,\Omega_2}&=-\frac{\Omega_1}{2\pi}\epsilon_{ilm}\delta_{\alpha m}\sum_{\kv,\kv'}\left(k'-k\right)^l
\gr_{\kv-\frac{\Qv}{2}}\ga_{\kv+\frac{\Qv}{2}}\left(\ga_{\kv+\frac{\qv}{2}}-\gr_{\kv-\frac{\qv}{2}}\right)\left(\ga_{\kv'+\frac{\Qv}{2}}-\gr_{\kv'-\frac{\Qv}{2}}\right),
\end{align*}
where $\rm{c.c}$ denotes the complex conjugate.
After expanding with respect to the wave vector $\qv$ and $\Qv$ (assuming $|\qv|l, |\Qv|l\ll1$), we obtain 
\begin{align*}
{\cal D}^{2(a)}_{\alpha i;\qv,\Qv,\Omega_1,\Omega_2}
&\simeq i\frac{2\eta}{3\pi} \Omega_1\epsilon_{i\alpha m}Q^m\sum_{\kv,\kv'}\ek |\gr_{\kv}|^2 |\gr_{\kv'}|^4,
\\
{\cal D}^{2(b+c)}_{\alpha i;\qv,\Qv,\Omega_1,\Omega_2}
&\simeq i\frac{2\eta}{3\pi}\Omega_{1}\epsilon_{i\alpha m}\left(q^m\sum_{\kv,\kv'}\ek|\gr_{\kv}|^4 |\gr_{\kv'}|^2+(\frac{Q-q}{2})^m\sum_{\kv,\kv'}\ek|\gr_{\kv}|^2 |\gr_{\kv'}|^4\right),
\\
{\cal D}^{2(d)}_{\alpha i;\qv,\Qv,\Omega_1,\Omega_2}
&\simeq \frac{i \eta}{3\pi}\Omega_{1}\epsilon_{i\alpha m}\left(q+3Q\right)^m\sum_{\kv, \kv'}\ek|\gr_{\kv}|^4|\gr_{\kv'}|^2.
\end{align*}
Using $
\sum_{\kv,\kv'}\ek |\gr_{\kv}|^2 |\gr_{\kv'}|^4=\sum_{\kv,\kv'}\ek|\gr_{\kv}|^4 |\gr_{\kv'}|^2=\frac{(N(0)\pi)^2\ef}{2\eta^3},
$
we obtain
\begin{align}
{\cal D}^{2(a)}_{\alpha i;\qv,\Qv,\Omega_1,\Omega_2}+{\cal D}^{2(b+c)}_{\alpha i;\qv,\Qv,\Omega_1,\Omega_2}={\cal D}^{2(d)}_{\alpha i;\qv,\Qv,\Omega_1,\Omega_2}=i\frac{N(0)^2\pi\ef}{6\eta^3}\Omega_1\epsilon_{i \alpha m}\left(q+3Q\right)^m.
\label{eq:D2}
\end{align}
We see that skew--scattering contribution $\left({\cal D}^{2(a)}_{\alpha i;\qv,\Qv,\Omega_1,\Omega_2}+{\cal D}^{2(b+c)}_{\alpha i;\qv,\Qv,\Omega_1,\Omega_2}\right)$ and the side jump contribution $\left({\cal D}^{2(d)}_{\alpha i;\qv,\Qv,\Omega_1,\Omega_2}\right)$ are same.
\subsubsection{Diffusive contribution}
We evaluate here the contribution which include the vertex corrections as shown in  Fig. 3(a)(b)(c). 
This contribution ${\cal D}^{(3)}_{\alpha i; \qv,\Qv, \Omega_1,\Omega_2}$ is given as
\begin{align}
{\cal D}^{(3)}_{\alpha i; \qv,\Qv, \Omega_1,\Omega_2}=&
\frac{\hbar^2}{m}\epsilon_{ilm}
\sum^n_{\nu=1}\gamma^\nu\sum_{\kv \kv'}k^\alpha  
\\\nonumber
&\times\int\frac{d\omega}{2\pi}
\left\{
\left(\Pi^{\nu-1}_{n=0}g_{\kv_{n}-\frac{\Qv}{2},\omega}\right)
\left[{\cal G}^{lm,3(a)}_{\qv,\Qv,\omega,\Omega_1,\Omega_2}+{\cal G}^{lm,3(b+c)}_{\qv,\Qv,\omega,\Omega_1,\Omega_2}\right]
\left(\Pi^{\nu}_{n=1}g_{\kv_{\alpha-n}+\frac{\Qv}{2},\omega+\tilde{\Omega}}\right)
\right\}
^{<},
\end{align}
where $\gamma=\frac{u_0^2\nimp}{V}$, $k_{0}\equiv k$ and
\begin{align*}
{\cal G}^{lm,3(a)}_{\qv,\Qv,\omega,\Omega_1,\Omega_2}=&\sum_{\kv_{\nu} \kv'_{\nu}}(k'_\nu-k_\nu)^l Q^m g_{\kv_\nu-\frac{\Qv}{2},\omega}g_{\kv'_\nu-\frac{\Qv}{2},\omega}g_{\kv'_\nu+\frac{\qv}{2},\omega+\Omega_{1}}g_{\kv'_\nu+\frac{\Qv}{2},\omega+\tilde{\Omega}}g_{\kv_\nu+\frac{\Qv}{2},\omega+\tilde{\Omega}},
\\
{\cal G}^{lm,3(b+c)}_{\qv,\Qv,\omega,\Omega_1,\Omega_2}=&\sum_{\kv_{\nu} \kv'_{\nu}}\left[2k'^l_\nu k^m_{\nu}+(k'_\nu-k_\nu)^l (\frac{Q+q}{2})^m\right]
\\
\times &g_{\kv_{\nu}-\frac{\Qv}{2},\omega}g_{\kv_{\nu}+\frac{\qv}{2},\omega+\Omega_1}g_{\kv'_{\nu}+\frac{\qv}{2},\omega+\Omega_{1}}g_{\kv'_{\nu}+\frac{\Qv}{2},\omega+\tilde{\Omega}}g_{\kv_{\nu}+\frac{\Qv}{2},\omega+\tilde{\Omega}}
\\
+&\sum_{\kv_{\nu} \kv'_{\nu}}\left[-2k'^l_{\nu} k^m_{\nu}+(k'_\nu-k_\nu)^l (\frac{Q-q}{2})^m\right]
\\
\times &g_{\kv_{\nu}-\frac{\Qv}{2},\omega}g_{\kv'_{\nu}-\frac{\Qv}{2},\omega}g_{\kv'_{\nu}+\frac{\qv}{2},\omega+\Omega_{1}}g_{\kv_{\nu}+\frac{\qv}{2},\omega+\Omega_{1}}g_{\kv_{\nu}+\frac{\Qv}{2},\omega+\tilde{\Omega}}.
\end{align*}
An infinite series over the index $\nu$ gives the vertex corrections shown diagrammatically in Fig. 3(d).
 After calculating the lesser component of the ${\cal D}^{(3)}_{\alpha i; \qv,\Qv, \Omega_1,\Omega_2}$,
 we obtain
 \begin{align*}
{\cal D}^{(3)}_{\alpha i; \qv,\Qv, \Omega_1,\Omega_2}=&\frac{\hbar^2}{m}\epsilon_{ilm}\sum^n_{\nu=1}\gamma^\nu\sum_{\kv \kv'}k^\alpha
\\
&\times\int\frac{d\omega}{2\pi}
\left\{
\left(\Pi^{\alpha-1}_{n=0}g_{\kv_{n}-\frac{\Qv}{2},\omega}\right)^{\rm r}
\left[{\cal G}^{lm,3(a),<}_{\qv,\Qv,\omega,\Omega_1,\Omega_2}+ {\cal G}^{lm,3(b+c), <}_{\qv,\Qv,\omega,\Omega_1,\Omega_2}\right]
\left(\Pi^{\alpha}_{n=1}g_{\kv_{\alpha-n}+\frac{\Qv}{2},\omega+\tilde{\Omega}}\right)^{\rm a}
\right\}.
\end{align*}
Here 
\begin{align*}
&{\cal G}^{lm, 3(a),<}_{\qv,\Qv, \Omega_1,\Omega_2}=-\delta(\omega)
\Omega_1\sum_{\kv_{\nu} \kv'_{\nu}}\left(k'_\nu-k_\nu\right)^l Q^m
 \left[
\gr_{\kv_{\nu}-\frac{\Qv}{2}}\gr_{\kv'_{\nu}-\frac{\Qv}{2}}\ga_{\kv'_{\nu}+\frac{\Qv}{2}}\ga_{\kv_{\nu}+\frac{\Qv}{2}}\left(\ga_{\kv'_{\nu}+\frac{\qv}{2}}-\gr_{\kv'_{\nu}-\frac{\qv}{2}}\right)
\right],
\\
&{\cal G}^{lm,3(b+c),<}_{\qv,\Qv, \Omega_1,\Omega_2}= -\delta(\omega)\Omega_1
 \sum_{\kv_{\nu},\kv_{\nu}'}
\left(-2k_{\nu}'^lk_{\nu}^m+\left(k_{\nu}'-k_{\nu}\right)^l\left(\frac{Q-q}{2}\right)^m \right)
\left[\gr_{\kv_{\nu}-\frac{\Qv}{2}}\gr_{\kv_{\nu}'-\frac{\Qv}{2}}\ga_{\kv_{\nu}'+\frac{\qv}{2}}\ga_{\kv_{\nu}+\frac{\qv}{2}}\ga_{\kv_{\nu}+\frac{\Qv}{2}} +{\rm c.c}\right] .
\end{align*}
Expanding ${\cal G}^{lm, 3(a),<}_{\qv,\Qv, \Omega_1,\Omega_2}$ and ${\cal G}^{lm, 3(b+c),<}_{\qv,\Qv, \Omega_1,\Omega_2}$ with respect to $\Qv$ and $\qv$, we obtain
\begin{align*}
{\cal G}^{lm, 3(a),<}_{\qv,\Qv, \Omega_1,\Omega_2}
\simeq-\frac{1}{3}Q^m q^l\delta(\omega)
\Omega_1  \sum_{\kv_{\nu} \kv'_{\nu}}\epsilon_{\kv'_{\nu}}|\gr_{\kv'_{\nu}}|^2\left[\left(\gr_{\kv'_{\nu}}\right)^2+\left(\ga_{\kv'_{\nu}}\right)^2\right]|\gr_{\kv_{\nu}}|^2,
\end{align*} 
and
\begin{align*}
{\cal G}^{lm, 3(b+c),<}_{\qv,\Qv, \Omega_1,\Omega_2} \simeq \frac{2}{3}Q^m q^l \delta(\omega)
\Omega_1  \eta^2\sum_{\kv_{\nu},\kv'_{\nu}}\epsilon_{\kv'_{\nu}}|\gr_{\kv'_{\nu}}|^4|\gr_{\kv_{\nu}}|^4.
 \end{align*}
Using $\sum_{\kv}\epsilon_{\kv}|g^{\rm r}_{\kv}|^2\left[(g^{\rm r}_{\kv})^2+(g^{\rm a}_{\kv})^2\right]\simeq-\frac{N(0)\pi}{2 \epsilon_F^2}(\frac{\epsilon_{F}}{\eta})^3,
\sum_{\kv} \epsilon_\kv |g^{\rm r}_{\kv}|^4\simeq\frac{N(0)\pi}{2\epsilon_F^2}(\frac{\epsilon_F}{\eta})^3,
\sum_{\kv} |g^{\rm r}_{\kv}|^4 \simeq\frac{N(0)\pi}{2\epsilon_F^3}
(\frac{\epsilon_F}{\eta})^3
$, we obtain 
\begin{align}
 {\cal G}^{lm, 3(a),<}_{\qv,\Qv, \Omega_1,\Omega_2}= {\cal G}^{lm, 3(b+c),<}_{\qv,\Qv, \Omega_1,\Omega_2}=\frac{1}{6}\delta(\omega) 
\Omega_1 Q^m q^l \frac{(N(0)\pi)^2\ef}{\eta^4},
\label{eq:g3}
\end{align}
and
\begin{align}
&{\cal D}^{(3)}_{\alpha,i;\qv,\Qv, \Omega_1,\Omega_2}
=&\frac{(N(0)\pi)^2\hbar\ef}{6\pi \eta^4}\Omega_1\epsilon_{ilm} Q^m q^l \sum^{\infty}_{\nu=1} \frac{\hbar}{m}\sum_{\{k_{n < \nu }\}}\gamma^\nu k^\alpha \left(\Pi^{\nu-1}_{n=0}g_{\kv_{n}-\frac{\Qv}{2}}\right)^{\rm r}\left(\Pi^{\nu}_{n=1}g_{\kv_{\alpha-n}+\frac{\Qv}{2},\tilde{\Omega}}\right)^{\rm a}.
\label{eq:DD-m}
\end{align}
 Summation over $\nu$ in eq. (\ref{eq:DD-m}) is carried out as (see Appendix),
\begin{align*}
\sum^{\infty}_{\nu=1} \frac{\hbar}{m}\sum_{\{k_{n < \nu }\}}\gamma^\nu k^\alpha (\Pi^{\nu-1}_{n=0}g_{\kv_{n}-\frac{\Qv}{2}})^{\rm r}(\Pi^{\nu}_{n=1}g_{\kv_{\nu-n}+\frac{\Qv}{2},\Omega_{1}+\Omega_{2}})^{\rm a}
\simeq\frac{iDQ^\alpha}{(i\tilde\Omega+DQ^2)\tau},
\end{align*}
and thus
\begin{align}
&{\cal D}^{(3)}_{\alpha i; \qv,\Qv, \Omega_1,\Omega_2}=\epsilon_{ilm}\frac{(N(0)\pi)^2 \ef}{3\pi \eta^3}\frac{iDQ^\alpha Q^m q^l}{i\tilde\Omega+DQ^2}.
\label{eq:D3}
\end{align}
Using the results eqs. (\ref{eq:D2}) and (\ref{eq:D3}), we finally obtain the charge current at second order in $\Jex$ $\left(\right.$eq. (\ref{eq:j2})$\left.\right)$ as 
\begin{align}
j_{\alpha}^{(2)}(\qv,\Qv,\Omega_1,\Omega_2)
=-\frac{\alpha_{\rm ISH} c \Jex}{2\eta}\epsilon_{ijk}\Omega_1
S^j_{\frac{\Qv+\qv}{2},\Omega1}S^k_{\frac{\Qv-\qv}{2},\Omega2}\left(\epsilon_{i\alpha m}\left(q+3Q\right)^m+\epsilon_{ilm}\frac{DQ^\alpha Q^m q^l}{i\tilde \Omega+DQ^2}\right).
\label{eq:resultj2}
\end{align}
\subsection{Charge density}
 \begin{figure} 
\begin{center}
\resizebox{8cm}{!}{\includegraphics[angle=0]{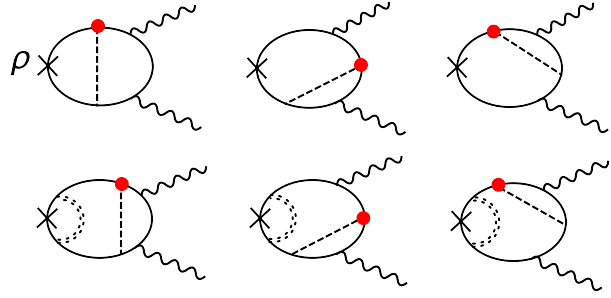}}
 \caption{Diagrammatic representation of the charge density.}
 \end{center}
  \end{figure}
  
We calculate the charge density in the similar way. The charge density is diagrammatically shown in Fig. 4, and is given as
\begin{align*}
\rho(\qv, \Qv,\Omega_1,\Omega_2)=i\frac{2e\hbar \Jex^2 \lamso u_{0}\nimp}{V^2}\epsilon_{ijk}\epsilon_{ilm}S^j_{\frac{\Qv+\qv}{2},\Omega_1}S^k_{\frac{\Qv-\qv}{2},\Omega_2}\left({\cal A}_{lm; \qv,\Qv, \Omega_1,\Omega_2}+ {\cal B}_{lm; \qv,\Qv, \Omega_1,\Omega_2}\right).
\end{align*}
Here,
\begin{align}
&{\cal A}_{lm; \qv,\Qv, \Omega_1,\Omega_2}=\int\frac{d\omega}{2\pi}
\left({\cal G}^{lm,3(a)}_{\qv,\Qv,\omega,\Omega_1,\Omega_2}+{\cal G}^{lm,3(b+c)}_{\qv,\Qv,\omega,\Omega_1,\Omega_2}\right)^{<},
\\
&{\cal B}_{lm; \qv,\Qv, \Omega_1,\Omega_2}=\sum^\infty_{\nu=1}\gamma^\nu\sum_{\{\kv_{n < \nu }\}}
\int\frac{d\omega}{2\pi}\left\{
\left(\Pi^{\nu-1}_{n=0}g_{\kv_{n}-\frac{\Qv}{2},\omega}\right)\left[{\cal G}^{lm,3(a)}_{\qv,\Qv,\omega,\Omega_1,\Omega_2}+{\cal G}^{lm,3(b+c)}_{\qv,\Qv,\omega,\Omega_1,\Omega_2}\right]\left(\Pi^{\nu}_{n=1}g_{\kv_{\nu-n}+\frac{\Qv}{2},\omega+\tilde{\Omega}}\right)\right\}^{<}.
\nonumber
\end{align}
By using the result of ${\cal G}^{lm,3(a),<}_{\qv,\Qv,\omega,\Omega_1,\Omega_2}$ and ${\cal G}^{lm,3(b+c),<}_{\qv,\Qv,\omega,\Omega_1,\Omega_2}$ (eq. (\ref{eq:g3})), we obtain
\begin{align*}
&{\cal A}_{lm; \qv,\Qv, \Omega_1,\Omega_2}\simeq\frac{N(0)^2\pi\ef}{6\eta^4}\Omega_1 Q^m q^l,
\\
&{\cal B}_{lm; \qv,\Qv, \Omega_1,\Omega_2}\simeq {\cal A}_{lm; \qv,\Qv, \Omega_1,\Omega_2} \tilde\Lambda^0_{\Qv,\tilde \Omega},
\end{align*}
where $\tilde \Lambda^0_{\Qv,\tilde \Omega}\equiv \sum^{\infty}_{\nu=1}\sum_{\{\kv_{n < \nu }\}}\gamma^\nu(\Pi^{\nu-1}_{n=0}g_{\kv_{n}-\frac{\Qv}{2},\omega})^{\rm r}(\Pi^{\nu}_{n=1}g_{\kv_{\nu-n}+\frac{\Qv}{2},\omega+\tilde{\Omega}}])^{\rm a}$. The details of calculation are presented in Appendix.
After some calculation, $\tilde \Lambda^0_{\Qv,\tilde \Omega}$ is obtained as
\begin{align*}
\tilde \Lambda^0_{\Qv,\tilde \Omega}
\simeq -1+\frac{1}{(DQ^2+i\tilde\Omega)\tau}.
\end{align*}
We thus obtain 
$
{\cal A}_{lm; \qv,\Qv, \Omega_1,\Omega_2}+{\cal B}_{lm; \qv,\Qv, \Omega_1,\Omega_2}
=\frac{N(0)^2\pi\ef}{6\eta^4}\Omega_1\frac{ Q^m q^l }{(i\tilde\Omega+DQ^2)\tau}.
$
Therefore the charge density is
\begin{align}
\rho(\qv,\Qv,\Omega_1,\Omega_2)=i\frac{\alpha\Jex}{2\eta}\epsilon_{ijk}\epsilon_{ilm}\Omega_1 S^j_{\frac{\Qv+\qv}{2},\Omega_1}S^k_{\frac{\Qv-\qv}{2},\Omega_2}\frac{Q^mq^l}{i\tilde \Omega+DQ^2}.
\label{eq:resultrho}
\end{align}
Comparing the second term of eq. (\ref{eq:resultj2}) (we define this term $\delta \tilde{j}_{\alpha}$) and eq. (\ref{eq:resultrho}), $\delta \tilde{j}_{\alpha}$ and the charge density $\rho$
are related as
\begin{align}
\delta \tilde{j}_{\alpha}(\qv,\Qv,\Omega_1,\Omega_2)
=iDQ^\alpha\rho(\qv,\Qv,\Omega_1,\Omega_2).
\label{eq:del2}
\end{align}
\section{Result and Discussion}
Summarizing the results (eqs. (\ref{eq:reslutj1}), (\ref{eq:resultj2}), (\ref{eq:resultrho}) and (\ref{eq:del2})),
we see that the electric current is written in the real space as 
\begin{align} 
j_\alpha = &\alpha_{\rm ISH} c \epsilon_{\alpha \beta \gamma}\left\{
\partial_\beta \dot S^\gamma + \frac{2J_{\rm{ex}}\tau}{\hbar}\left[\partial_\beta\left(\Sv \times \dot{\Sv}\right)^\gamma + \left(\Sv\times \partial_\beta \dot{\Sv}\right)^\gamma\right]\right\} - D\partial_\alpha \rho. 
\label{eq:j1} 
\end{align}
This is the main result of this paper.
The first term in the curly bracket is the local contribution arising from the processes illustrated in Figs. 1 and 2. The last term $\partial_\alpha \rho$ is obtained by evaluating the diffusion ladder (vertex corrections) illustrated in Fig. 4.
Including these vertex corrections is equivalent to imposing the gauge invariance by using the Ward-Takahashi identity \cite{Langer,Engelsberg,Mahan}.
The expression of the charge density (eq. (\ref{eq:resultrho})) is given in real space as
\begin{align*} 
\rho(\xv,t) = \frac{2\alpha_{\rm ISH} c\Jex \tau}{\hbar} \epsilon_{\alpha \beta \gamma}\epsilon_{\alpha j k}\int dt_1 \int d\xv_1\zeta(\xv-\xv_1,t-t_1){\partial_k}\dot{S}^\beta_{\xv_1,t_1}{\partial_j}S^\gamma_{\xv_1,t_1}, 
\end{align*}
where
 \begin{align*}
 \zeta(\xv,t)=\frac{1}{2\pi V}\sum_{\Qv,\tilde{\Omega}}\frac{e^{i(\tilde{\Omega}t-\Qv\cdot\xv)}}{DQ^2+i\tilde{\Omega}},
 \end{align*}
  represents the retarded diffusion propagator satisfying $(\partial_t-D\nabla^2)\zeta(\xv,t)=\delta(\xv)\delta(t)$. 
The consistency of our calculation is seen from the charge and current densities obtained above satisfying the charge conservation law, $\dot{\rho}+\nabla\cdot\jv=0$, neglecting higher order derivatives.

The direct (local) part of the current in eq. (\ref{eq:j1}) indicates that charge current is induced if magnetization is dynamic, and in particular, if the
magnetization has energy dissipation, $\Sv\times \dot{\Sv}$. The current arising from the damping was found also in a Rashba spin--orbit system \cite{Ohe07,Takeuchi08}, 
but there was no contribution proportional to $\dot{\Sv}$ in the uniform Rashba case.

An interesting observation is that the direct contribution is expressed by the spin current pumped in the system. Let us define the spin current by the normal part (i.e.,
without spin--orbit interaction) as $j_{{\rm s},\beta}^{\gamma }=-i\frac{e\hbar}{2m}\langle\psi^{\dagger}\sigma^\gamma \overleftrightarrow{\partial_{\beta}}\psi\rangle$.
In the context of the spin Hall effect, proper definition of the spin current has been argued intensively \cite{Sun05,Shi06}.
In the present issue of the inverse spin Hall effect, in contrast, for the present issue of the inverse spin Hall effect, only the normal part of the spin current needs to be considered, since other terms related with the spin relaxation by the spin--orbit interaction give only higher order contribution in $\lamso$.
 The result of the spin current is given by \cite{Takeuchi08}
\begin{align}
j_{{\rm s},\beta}^{\gamma}=c\left\{\partial_\beta\dot S^\gamma+\frac{2\Jex \tau}{\hbar}\left[\partial_\beta\left(\Sv \times \dot{\Sv}\right)^\gamma + \left(\Sv\times \partial_\beta\dot{\Sv}\right)^\gamma\right]\right\}.
\label{eq:spincurrent}
\end{align}
 We see the direct part of the charge current, eq. (\ref{eq:j1}), is simply written in terms of the spin current, giving eq. (\ref{eq:jfull}). 
The equation for the spin current created by the spin dynamics, eq. (\ref{eq:spincurrent}), indicates that spin current is pumped when 1) there is either spin dynamics ($\dot{\Sv}$)
or spin damping $(\Sv\times \dot{\Sv})$ and 2) the spin has spatial modulation (i.e., there is a finite spatial derivative). Applying the result to the junction system with the
interface chosen as $z=0$, perpendicular to the $z$ axis, we may replace $\partial_z\Sv$ by $\delta(z)\Sv$. The spin current in this case flows at the interface
and reduces to $j_{{\rm s}, z}^{\gamma }\mid_{z=0}\simeq c\delta(0)\{\dot{S}^\gamma + \frac{6\Jex\tau}{\hbar} (\Sv\times\dot{\Sv})^\gamma\}$ at the interface, in
agreement with phenomenological spin pumping theory \cite{Tserkovnyak02}. 

Our weak coupling result seems very different from that for the strong coupling regime \cite{Stern92,Duine07}. In ref. \citen{Duine07}, the spin--orbit contribution to the
pumped current was argued to be the scalar product of localized spins, $j_\alpha \sim \beta_{\rm sr}({\dot \Sv}\cdot\partial_\alpha \Sv)$, while in the weak coupling
regime, only the vector product appears in the expression for the current. This situation is different from linear response transport properties such as the anomalous Hall effect owing to the
spin structure; the weak and strong coupling limits in the case of the anomalous Hall effect agree when the spin structure slowly varies \cite{Onoda04}. 

The direct part of the current (eq. (\ref{eq:j1})) indicates that the term 
\begin{align}
c \epsilon_{\alpha \beta \gamma}\left\{\partial_\beta\dot{S}^\gamma +\frac{2J_{\rm ex}\tau}{\hbar}
\left[\partial_\beta\left(\Sv\times\dot{\Sv}\right)+\Sv\times\partial_\beta\dot{\Sv}\right]^\gamma \right\}\equiv \sigma E_\alpha^{\rm eff}
\end{align}
acts as the effective field that drives the charge
current, where $\sigma$ is the Boltzmann conductivity. Volovik pointed out that such a field exists in a strong coupling regime in the absence of the spin--orbit
interaction \cite{Volovik}. 
Volovik's electric field is written in terms of the spin variable as $E_\alpha^{\rm (V)}=\frac{e^2}{h}\Sv\cdot(\dot{\Sv}\times \partial_\alpha \Sv)$, which is the Berry phase in space and time. This field is written by taking $\Ev^{\rm (V)}=\partial_t \Av^{\rm (V)}-\nabla A_0^{\rm (V)}$ as U(1), where $A_\mu^{\rm (V)}\equiv -\partial_\mu\phi \cos\theta$ ($\theta$ and $\phi$ are polar angles of the spin $\Sv$ and $\mu=0, x,y,z$). This U(1) gauge field arises by projecting the SU(2) gauge field of the spin onto the polarization direction. The spin gauge field in the
original SU(2) space is given by $\Av_\mu=\frac{1}{2}(\nv\times \partial_\mu\nv) + \frac{1}{2} A_\mu^{\rm (V)}\nv$ ($\nv\equiv \Sv/S$) \cite{Volovik,Tatara} and thus 
Volovik's gauge field is $A_\mu^{\rm (V)}=2(\nv\cdot\Av_\mu)$. In the absence of the spin--orbit interaction and in the strong coupling limit, only this component couples
to the real charge and becomes physical. This is understandable since only the component along the polarization direction is essential  for this limit. 
In contrast, the effective electric field for the weak coupling limit with the spin--orbit interaction arises from the component of the SU(2) gauge field perpendicular to the adiabatic component.
Namely, defining
$\Av_\mu^{\perp}\equiv 2\Av_\mu- A_\mu^{\rm (V)}\nv=\nv \times \partial_\mu \nv
$, the effective field is written as 
\begin{align}
E_\alpha^{\rm (eff)}=\frac{c S}{\sigma}\epsilon_{\alpha \beta \gamma} \left\{
-\partial_\beta \left(\nv \times \Av_0^{\perp}\right)^\gamma
+\frac{2\Jex S \tau}{\hbar}\left[\partial_\beta \Av_0^{\perp} + \partial_t \Av_\beta^{\perp} - \left(\dot{\nv}\times\partial_\beta\nv\right)\right]^\gamma\right\}.
\end{align} 
(We note that both $\Av_\mu^{\perp}$ and 
$\nv\times\Av_\mu^{\perp}$ are  perpendicular to $\nv$ and to the adiabatic component. 
 The last term is not changed since not all terms can be written using the gauge field in the perturbative regime). Therefore, the dynamic inverse spin Hall effect is due to a
skewed projection of the SU(2) gauge field onto the physical U(1) field induced by the relativistic spin--orbit interaction. Owing to the skewed projection, the spatial
coordinates and the spin coordinates are mixed (by the anti-symmetric tensor $\epsilon_{\alpha \beta\gamma}$) in the effective electric field. 
Shibata and Kohno calculated the charge current induced by magnetization dynamics for the strong coupling limit and in the presence of the spin--orbit interaction \cite{Shibata}.
They showed the induced Hall current was perpendicular to Volovik's electric field $\jv \propto \lamso (\nv\times \Ev^{\rm (V)})$.
\begin{figure} 
\begin{center}
\resizebox{5cm}{!}{\includegraphics[angle=0]{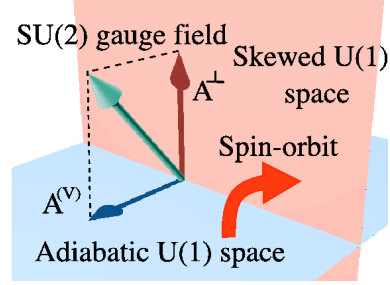}}
 \caption{(Color online) Schematic illustration of the effective electric (U(1)) field arising from the projection of the SU(2) spin gauge field. For the strong coupling limit (adiabatic limit), 
 the U(1) gauge field $A_\mu^{(V)}$ is that projected along the spin direction, while the dynamic inverse spin Hall effect is driven by the component perpendicular to the adiabatic component $A_\mu^{\perp}$.}
 \end{center}
  \end{figure}

Let us estimate the magnitude of the spin current and expected charge current. 
Assuming the magnetization precesses at the interface with an angle $\theta$ and frequency $f$,
the magnitude of the spin current density is evaluated from eq. (\ref{eq:spincurrent}) as
\begin{equation*}
 |j_{\rm s}|=\frac{2 e\kf^2}{3\pi^2}\biggl(1+6\frac{\Jex\tau}{\hbar}\sin\theta\biggr)\left(\frac{\Jex\ef\tau^2}{\hbar^2}\right) f \; {\rm
A/m}^2.
\end{equation*}
For a disordered metal with
$\frac{\hbar}{\ef \tau}\simeq 0.1$ coupled weakly to the magnetization ($\frac{\Jex}{\ef}=0.01$), the spin current density is estimated to be $|j_{\rm s}|=1\times
10^{10}$ A/m$^2$ for a frequency of $f=10$ GHz \cite{Saitoh06} and $\kf=1\times 10^{10}$ $\rm 1/m$ \cite{Takahashi08} if $\theta\sim O(0.1)$.
We estimate the spin--charge conversion efficiency as $\alpha_{\rm ISH}=1\times (10^{-3}\sim 10^{-2})$ 
using the dimensionless spin--orbit coupling $\kf^2\lamso =0.01\sim 0.1$ for various materials \cite{Takahashi08}. We therefore expect the magnitude of the output charge current to be $|j|\simeq 1\times
(10^7\sim 10^8)$ $\rm A/ m^2$.
\section{Conclusions}
In conclusion, we presented gauge-invariant theory for the dynamic inverse spin Hall effect satisfying charge conservation.
Assuming metallic systems, the case of symmetric spin--orbit interaction due to random impurities was considered.
We showed that charge current is induced directly by the effective electric field, which is proportional to the local spin current, but not by the diffusive process.
The effective field is explained as a skewed U(1) projection owing to the relativistic spin--orbit interaction.
\section{Acknowledgment}
The authors thank E. Saitoh, J. Shibata, H. Kohno, S. Murakami and S. Takahashi for valuable discussions. Two of the authors (K.H. and A.T.) would like to express their gratitude for support provided by the Ministry of Education, Culture, Sports, Science and Technology (MEXT, Japan) program "Support Program for Improving Graduate School Education". This work was supported by a Grant-in-Aid for Scientific Research in Priority Areas, "Creation and control of spin current" (1948027), the Kurata Memorial Hitachi Science and Technology Foundation and the Sumitomo Foundation. 

\appendix
\section{Derivation of the diffusion factor}
We present, in this section, details of the calculation of the infinite series summation due to the vertex corrections \cite{Bruus,Rammer}. In the calculation, we see 
\begin{align*}
\tilde{\Lambda}^\alpha_{\Qv,\tilde \Omega}\equiv \sum^{\infty}_{\nu=1} \frac{\hbar}{m}\sum_{\kv_{n < \nu }}\gamma^{\nu} k^{\alpha} (\Pi^{\nu-1}_{n=0}g_{\kv_{n}-\frac{\Qv}{2}})^{\rm r}(\Pi^{\nu}_{n=1}
g_{\kv_{\nu-n}+\frac{\Qv}{2},\tilde{\Omega}})^{\rm a},
\\
\tilde \Lambda^0_{\Qv,\tilde \Omega}\equiv \sum^{\infty}_{\nu=1} \sum_{\kv_{n < \nu }}\gamma^\nu (\Pi^{\nu-1}_{n=0}g_{\kv_{n}-\frac{\Qv}{2}})^{\rm r}(\Pi^{\nu}_{n=1}g_{\kv_{\nu-n}+\frac{\Qv}{2},\tilde{\Omega}})^{\rm a}.
\end{align*}
Above equations are rewritten as
\begin{align}
\tilde{\Lambda}^\alpha_{\Qv,\tilde \Omega}=&\Lambda^\alpha_{\Qv,\tilde \Omega}\{1+\Lambda^0_{\Qv,\tilde \Omega}+(\Lambda^0_{\Qv,\tilde \Omega})^2+(\Lambda^0_{\Qv,\tilde \Omega})^3+\cdots\}
=\Lambda^\alpha_{\Qv,\tilde \Omega}\{1+\tilde \Lambda^0_{\Qv,\tilde \Omega}\},
\nonumber\\
\tilde{\Lambda}^0_{\Qv,\tilde \Omega} \equiv &\Lambda^0_{\Qv,\tilde \Omega}+(\Lambda^0_{\Qv,\tilde \Omega})^2+(\Lambda^0_{\Qv,\tilde \Omega})^3+\cdots
=\Lambda^0_{\Qv,\tilde \Omega}\{1+\tilde \Lambda^0_{\Qv,\tilde \Omega}\},
\label{eq:lambda2}
\end{align}
where we define
\begin{align*}
\Lambda^{\alpha}_{\Qv,\tilde \Omega} \equiv &\frac{\hbar}{m}\gamma\sum_{\kv}k^\alpha g_{\kv-\frac{\Qv}{2}}^{\rm r}g_{\kv+\frac{\Qv}{2},\tilde{\Omega}}^{\rm a},
\\
\Lambda^0_{\Qv,\tilde \Omega} \equiv &\gamma\sum_{\kv} g_{\kv-\frac{\Qv}{2}}^{\rm r}g_{\kv+\frac{\Qv}{2},\tilde{\Omega}}^{\rm a}.
\end{align*}
Expanding $\Lambda^{\alpha}_{\Qv,\tilde \Omega}$ and $\Lambda^0_{\Qv,\tilde \Omega} $ with respect to $\Qv,\tilde{\Omega}$ in
slowly varying limit, we obtain as
\begin{align*}
\Lambda^{\alpha}_{\Qv,\tilde \Omega} = &iDQ^\alpha,\\
\Lambda^0_{\Qv,\tilde \Omega} =&1-(DQ^2+i \tilde\Omega)\tau.
\end{align*}
Since we obtain $\tilde {\Lambda}^0_{\Qv,\tilde \Omega} =\frac{\Lambda^0_{\Qv} }{1-\Lambda^0_{\Qv} }$ from eq. (\ref{eq:lambda2}), $\tilde {\Lambda}^{\alpha}_{\Qv}$ and $\tilde {\Lambda}^0_{\Qv}$
are given as
\begin{align*}
\tilde {\Lambda}^{\alpha}_{\Qv} =&\frac{iDQ^\alpha}{(DQ^2+i\tilde \Omega)\tau},\\
\tilde {\Lambda}^0_{\Qv} =&-1+\frac{1}{(DQ^2+i\tilde \Omega)\tau}.
\end{align*}


\begin{thebibliography}{99}

\bibitem{Murakami} S. Murakami, N. Nagaosa, and S.-C. Zhang: Science {\bf 301} (2003) 1348.

\bibitem{Shinova} J. Sinova, D. Culcer, Q. Niu, N.A. Sinitsyn, T. Jungwirth, and A.H. MacDonald: Phys. Rev. Lett. {\bf 92} (2004) 126603.

\bibitem{Kato} Y.K. Kato, R.C. Myers, A.C. Gossard, and D.D Awshalom: Science {\bf 306} (2004) 1910.

\bibitem{JW} J. Wunderlich, B. Kaetner, J. Sinova, and T. Jungwirth: Phys. Rev. Lett. {\bf 94} (2005) 047204.

\bibitem{Saitoh06} E. Saitoh, M. Ueda, H. Miyajima, and G. Tatara: Appl. Phys. Lett. {\bf 88} (2006) 182509.

\bibitem{Takahashi02} S. Takahashi and S. Maekawa: Phys. Rev. Lett. {\bf 88} (2002) 116601.

\bibitem{Takahashi08} S. Takahashi and S. Maekawa: J. Phys. Soc. Jpn. {\bf77} (2008) 031009.

\bibitem{Valenzuela06} S.O. Valenzuela and M. Tinkham: Nature {\bf 442} (2006) 176.

\bibitem{Kimura07} T. Kimura Y. Otani, T. Sato, S. Takahashi, and S. Maekawa: Phys. Rev. Lett. {\bf 98} (2007) 156601.

\bibitem{Seki} T. Seki, Y. Hasegawa, S. Mitani, S. Takahashi, H. Imamura, S. Maekawa: Nat. Mater.  {\bf 7} (2008) 125.

\bibitem{Ando} K. Ando, Y. Kajiwara, S. Takahashi, S. Maekawa, K. Takemoto, M. Takatsu, and E. Saitoh: Phys. Rev. B {\bf 78} (2008) 014413.

\bibitem{Brataas00} A. Brataas, Y.V. Nazarov, and G.E.W. Bauer: Phys. Rev. Lett. {\bf 84} (2000) 2481.

\bibitem{Tserkovnyak02} Y. Tserkovnyak, A. Brataas, and G.E.W. Bauer: Phys. Rev. Lett. {\bf 88} (2002) 117601.

\bibitem{Wang06} X. Wang, G. E. W. Bauer, B. J. van Wees, A. Brataas and Y. Tserkovnyak: Phys. Rev. Lett. {\bf 97} (2006) 216602.

\bibitem{Ohe07} J.-I. Ohe, A. Takeuchi, and G. Tatara: Phys. Rev. Lett. {\bf 99} (2007) 266603.

\bibitem{Takeuchi08} A. Takeuchi and G. Tatara: J. Phys. Soc. Jpn. {\bf 77} (2008) 074701.

\bibitem{Ast07} C.R. Ast, J. Henk, A. Ernst, L. Moreschini, M.C. Falub, D. Pacil\'e, P. Bruno, K. Kern, and M. Grioni: Phys. Rev. Lett. {\bf 98} (2007) 186807.

\bibitem{Dugaev1} V.K. Dugaev, A. Cr\'epieux, and P. Bruno: Phys. Rev. B {\bf 64} (2001) 104411.

\bibitem{WK} W.-K. Tse and S.D. Sarma: Phys. Rev. Lett. {\bf 96} (2006) 056601.

\bibitem{KAM} K.A. Muttalib and P. W\"olfle: Phys. Rev. B {\bf 76} (2007) 214415.

\bibitem{Hung} H. Hung, A.P. Jauho, {\it Quantum Kinetics in Transport and Optics of Semi-conductors} (Springer-Verlag, 1998).

\bibitem{Langer} J.S. Langer: Phys. Rev. {\bf 124} (1961) 1003.

\bibitem{Engelsberg} S. Engelsberg and J.R. Schrieffer: Phys. Rev. {\bf 131} (1963) 993.

\bibitem{Mahan} G.D. Mahan: {\it Many-Particle Physics} (Plenum Press, New York, 1990)

\bibitem{Sun05} Q.-F. Sun  and X.C. Xie: Phys. Rev. B {\bf 72} (2005) 245305.

\bibitem{Shi06} J. Shi, P. Zhang, D. Xiao, and Q. Niu: Phys. Rev. Lett. {\bf 96} (2006) 076604.

\bibitem{Stern92} A. Stern: Phys. Rev. Lett. {\bf 68} (1992) 1022.

\bibitem{Duine07} R.A. Duine: Phys. Rev. B {\bf 77} (2008) 014409.

\bibitem{Onoda04} M. Onoda, G. Tatara, and N. Nagaosa: J. Phys. Soc. Jpn. {\bf 73} (2004) 2624.

\bibitem{Volovik} G.E. Volovik: J. Phys. C {\bf 20} (1987) L83.

\bibitem{Tatara} G. Tatara, H. Kohno, and J. Shibata: Phys. Rep. {\bf 468} (2008) 213-301.

\bibitem{Shibata} J. Shibata and H. Kohno: Phys. Rev. Lett. {\bf 102} (2009) 086603.

\bibitem{Bruus} H. Bruus and K. Flensberg: {\it Many-Body Quantum Theory in Condensed Matter Physics} (Oxford University Press, 2004)

\bibitem{Rammer} J. Rammer: {\it Quantum Field Theory of Non-equiliburium States} (Cambridge, 2007)

\end{thebibliography}
\end{document}